\documentclass[twocolumn,pra,showpacs,superscriptaddress]{revtex4}
\usepackage{amssymb}
\usepackage{amsmath}
\usepackage{graphicx}
\usepackage{subfigure}
\usepackage{natbib}
\usepackage{epsfig}
\usepackage{amsfonts}
\usepackage{mathrsfs}
\usepackage{ulem}
\usepackage{color}
\usepackage[toc,page,title,titletoc,header]{appendix}
\begin{document}
\title{Strongly interacting one-dimensional quantum gas mixtures with weak p-wave interactions}
\author{Haiping Hu}
\affiliation{Beijing National
Laboratory for Condensed Matter Physics, Institute of Physics,
Chinese Academy of Sciences, Beijing 100190, China}

\author{Lei Pan}
\affiliation{Beijing National
Laboratory for Condensed Matter Physics, Institute of Physics,
Chinese Academy of Sciences, Beijing 100190, China}

\author{Shu Chen}
\thanks{schen@aphy.iphy.ac.cn}
\affiliation{Beijing National
Laboratory for Condensed Matter Physics, Institute of Physics,
Chinese Academy of Sciences, Beijing 100190, China}
\affiliation{Collaborative Innovation Center of Quantum Matter, Beijing, China}

\begin{abstract}
We study one-dimensional strongly interacting quantum gas mixtures, including both the Bose-Fermi and spin-$1/2$ Fermi-Fermi mixtures, with weak p-wave interactions between intra-component fermions, and demonstrate that the weak p-wave interaction can not be omitted in the strongly interacting regime where the strength of p-wave interactions is comparable with the inverse of the strength of strongly repulsive s-wave interactions.
While the total density distribution is not sensitive to the weak p-wave interaction, we find that the p-wave interaction plays an important role in determining the species-dependent (or spin-dependent) density distributions and produces significant physical effects on the low-energy spin dynamics.
We also derive effective spin-exchange models for strongly interacting quantum gas mixtures with weak p-wave interactions, which indicate that a quantum phase transition from anti-ferromagnetic state to ferromagnetic state can be induced by tuning the relative strengths of intra-component and inter-component interactions.

\end{abstract}


\maketitle


\section{Introduction}
Strongly interacting one-dimensional (1D) quantum gases have attracted intensive studies in recent years due to the experimental progress in trapping and manipulating cold atomic systems \cite{RMP-Bose,RMP-Guan,Paredes,Kinoshita,Olshanii,Gorlitz,Moritz,Haller}. Particularly, tunability of interaction strengths between atoms has enabled
us to access the entire interaction parameter regime \cite{Paredes,Kinoshita,Olshanii,Gorlitz,Moritz}, which leads to the realization of Tonks-Girardeau (TG) gas \cite{Paredes,Kinoshita} and super-Tonks-Girardeau (STG) gas \cite{Haller}. Moreover, recent experiments on few-particle atomic systems \cite{Jochim1,Jochim2,Jochim3} with the controllability of precise atom numbers and interaction strengths open access to experimentally studying strongly interacting few-body physics. These symbolic experimental progresses stimulate the renewed theoretical interests in the study of TG gases \cite{Girardeau}. The Bose-Fermi mapping, which was originally proposed to solve the single-component Bose gas with infinite repulsion \cite{Girardeau}, has been generalized to deal with the multicomponent TG systems \cite{Girardeau07,Fang,Deuretzbacher,Guan09}, including Bose-Bose mixtures \cite{Girardeau07,Deuretzbacher}, spin-1/2 Fermi gases \cite{Girardeau07,Guan09}, and Bose-Fermi mixtures \cite{Girardeau07,Fang}.  A significant difference from the single-component TG gas is that the ground state of the multi-component TG gas is highly degenerate with the degree of degeneracy given by the number of configurations of different components \cite{Girardeau07,Fang,Deuretzbacher,Guan09}. Slightly away from the TG limit, the degeneracy of the ground states is lifted \cite{Guan09,Fang,Guan10,Cui,Harshman}, and various perturbation methods within the degenerate subspace can be constructed by treating the inverse of the interaction strength as a small parameter \cite{Guan-thesis,Zinner14,Zinner-SR,PuHan,Deuretzbacher14,Zinner15,hu,Levinsen,Levinsen2,cuixiaoling}.
Particularly, the effective spin-exchange Hamiltonian describing the spin dynamics of the multi-component system in the strongly repulsive regime has been derived \cite{PuHan,Deuretzbacher14,Zinner15,Levinsen,Levinsen2,cuixiaoling,Zinner-SR}. Numerical studies of multi-component few-particle systems in a wide regime of interaction parameter have also been carried out by different groups \cite{Hao,Blume,Schmelcher,Lewenstein,Zinner-EPL}. It is worth indicating that the experimental realization of antiferromagnetic (AFM) Heisenberg spin chain consisting of strongly interacting few cold atoms in a 1D trap has been reported very recently by Murmann et. al. \cite{Murmann}. Their experimental results are found to be in good agreement with the effective spin-exchange model for fermionized particles. 

In most of the previous theoretical works on multi-component quantum gases, only the s-wave interaction is considered, as the p-wave interaction is very weak in comparison with the s-wave interaction. However, in principle the p-wave interaction can be greatly enhanced by the Feshbach resonances \cite{Regal,Blume04,PengSG}. And in some cases, for example, for a spin-polarized fermionic gas, the p-wave interaction plays a  dominant role as the s-wave scattering is forbidden due to the Pauli exclusion principle. For a 1D polarized fermionic gas with p-wave interactions, some theoretical works have unveiled the existence of a Bose-Fermi duality between the 1D spinless p-wave fermionic system and a bosonic system with the reversed role of strong and weak couplings \cite{Cheon,Girardeau05,Girardeau06,Girardeau04,Granger,Grosse,Chen07}. Such studies inspire us to realize that the weak p-wave interaction may also play important role in 1D strongly interacting Bose-Fermi and Fermi-Fermi mixtures when the strength of p-wave interaction is comparable with the inverse of the strength of s-wave interaction. Therefore, it is interesting to study the quantum gas mixtures in the presence of both s-wave and p-wave interactions and explore the important physical effects of p-wave interactions.

In this work, by studying both the Bose-Fermi mixtures and Fermi-Fermi mixtures, we shall show that the weak p-wave interaction indeed plays an important role in determining the spin-dependent (species-dependent) properties of the quantum gas mixtures when the strength of p-wave interaction is comparable with the inverse of the strength of strong s-wave interaction. 
To understand this, we first discuss the energy expansion of the system close to its TG limit, which clearly indicates that the contribution from the p-wave interaction is comparable with the s-wave interaction. We then study the effect of p-wave interaction on ground state density distributions of the quantum gas mixtures by variational perturbation theory within the degenerate ground space of the systems in the TG limit. Our results indicate that the species-dependent density distributions highly rely on the parameter ratio of strengths of p-wave interaction and s-wave interaction, whereas the total density distribution is not sensitive to the anisotropic parameter. Finally, we demonstrate that the low-energy spin dynamics of the system can be described by an effective spin-exchange model and discuss quantum phase transition induced by tuning the anisotropic parameter of interactions between different components.  Our results unveil the important role of the weak p-wave interaction for 1D quantum gas mixtures with strong s-wave interaction
and demonstrate that the interplay between p-wave and s-wave interactions can produce quantum phase transition between phases with different magnetisms. In view of recent progress in experiment on studying quantum magnetism of few-atom systems \cite{Murmann}, our theoretical work may motivate the experimental exploration of quantum magnetism phase transition induced by the interplay between p-wave and s-wave interactions in a trap system without an underlying
lattice.

The structure of our paper is organized as follows. We shall first present a detailed study of strongly interacting two-component Bose-Fermi mixtures with weak p-wave interaction between intra-component fermions in section II, which includes three subsections. In the first subsection, we introduce the model, and then derive the universal energy relation in the presence of p-wave interaction. In the second subsection, a variational perturbation method is introduced and applied to calculate an example few-body system composed of two bosons and two fermions. In the third subsection, an effective Hamiltonian describing the spin dynamics in the strongly interacting limit is derived. From the effective Hamiltonian, we can see the important role of the p-wave interactions, which may induce phase transition between phases with different spin configurations. We then generalize our method to deal with two-component Fermi-Fermi mixtures with weak p-wave interactions between intra-component fermions in section III. Due to the similarity of the method to the one in the previous section, we shall omit the details and only present the main results. A summary is given in section IV.

\section{Bose-Fermi mixture with weak p-wave interactions}
\subsection{Model and energy expansion close to the TG limit}
We consider the 1D Bose-Fermi mixtures with equal masses described by the following Hamiltonian
\begin{eqnarray}
H&=& \sum_{i} [- \frac{\hbar^2}{2 m}\frac{\partial^2}{\partial x_{i}^2}+V(x_{i})]\notag\\\notag
 &+& g_{bb} \sum_{i<j} \delta(x_{i}-x_{j}) \delta^b_{\sigma_i,\sigma_j} + g_{bf}\sum_{i<j}\delta(x_{i}-x_{j})\delta_{\sigma_i,-\sigma_j}\\
  &+& g_{ff}\sum_{i<j}V_p\left( x_i -x_j \right) \delta^f_{\sigma_i,\sigma_j}, \label{H-contact}
\end{eqnarray}
where $\sigma_i=b,~f$ represent the bosonic and fermionic components, respectively, $\delta^{b}_{\sigma_i, \sigma_j}= 1$ only if $\sigma_i$=$\sigma_j$=$b$, $\delta^{f}_{\sigma_i, \sigma_j}= 1$ only if $\sigma_f$=$\sigma_j$=$f$,
$\delta_{\sigma_i, - \sigma_j}=1$ when ($\sigma_i=b$,  $\sigma_j=f$) or ($\sigma_i=f$,  $\sigma_j=b$), and the summation to the spin index is assumed. The trap potential $V(x)$ takes the same form for the bosons and fermions. The particle numbers for bosons and fermions are $N_b$ and $N_f$, respectively. Here $g_{bb}$ denotes the effective s-wave interaction parameter between intra-component bosons, and $g_{bf}$ denotes the effective interaction parameter between inter-component bosons and fermions. The p-wave interaction between intra-component fermions is described by the pseudo-potential
\begin{equation}
V_p\left( x_i -x_j \right)= (\frac{\partial}{\partial x_i}-\frac{\partial}{\partial x_j})\delta(x_i-x_j)(\frac{\partial}{\partial x_i}-\frac{\partial}{\partial x_j})
\end{equation}
with $g_{ff}=- \hbar ^2a_{1D}^F/2 m$ denoting the effective interaction parameter. We note that the $p$-wave
scattering of two spin-polarized fermions in a tightly confined
waveguide has been shown to be well described by the contact condition
\cite{Blume04,Girardeau04}
\begin{eqnarray}
\Psi\left( x_i-x_j=0^{+}\right)  &=&-\Psi\left( x_i-x_j=0^{-}\right)
\nonumber \\
&=&-a_{1D}^F\frac \partial {\partial x}\Psi _F\left( x_i=x_j\pm 0\right) ,
\end{eqnarray}
with the effective 1D scattering length given by
$
a_{1D}^F=\frac{3a_p^3}{l_{\perp }^2}\left[ 1+\frac{3\zeta (3/2)}{2\sqrt{2}%
\pi }\left( \frac{a_p}{l_{\perp }}\right) ^3\right] ^{-1}
$, where $a_p$ is the $p$-wave scattering
length and $l_{\perp }=\sqrt{\hbar /m\omega _{\perp }}$ the transverse
oscillator length \cite{Girardeau04}. It is easy to check that the contact condition can be
reproduced by using the given pseudo-potential $V_p\left( r \right)$ \cite{Grosse,Girardeau04}.
For the convenience of calculation, in the following we take $\hbar=1$ and $m=1$ and consider the dimensionless Hamiltonian
\begin{eqnarray}
H&=& \sum_{i} [- \frac{1}{2} \frac{\partial^2}{\partial x_{i}^2}+V(x_{i})]\notag\\
 &+& g_{bb} \sum_{i<j} \delta(x_{i}-x_{j}) \delta^b_{\sigma_i,\sigma_j} + g_{bf} \sum_{i<j}\delta(x_{i}-x_{j})\delta_{\sigma_i,-\sigma_j}\notag\\
  &+& g_{ff}\sum_{i<j}(\frac{\partial}{\partial x_i}-\frac{\partial}{\partial x_j})\delta(x_i-x_j)(\frac{\partial}{\partial x_i}-\frac{\partial}{\partial x_j})\delta^f_{\sigma_i,\sigma_j}, ~~~~ \label{H-BF}
\end{eqnarray}
where $g_{bb}$, $g_{bf}$ and $g_{ff}$ can be regarded as rescaled dimensionless interaction parameters.

When $g_{ff}=0$, the model reduces to the 1D interacting Bose-Fermi model, which is well studied in past decades \cite{Fukuhara,Das,Cazalilla,Mathey,Lai,Imambekov,Yin,Batchelor}. In the TG limit with $g_{bb} \rightarrow \infty$, $g_{bf} \rightarrow \infty$ and $g_{ff}=0$,  the Bose-Fermi mixture can be mapped to a polarized Fermi system by a generalized Bose-Fermi mapping \cite{Girardeau07, Fang}, with the many-body wavefunction given by
\begin{eqnarray}
\Psi(x_1,x_2,...,x_N)&=&A\psi_A(x_1,x_2,...,x_N)
\end{eqnarray}
where
\begin{eqnarray}
\psi_A(x_1,x_2,...,x_N)=\frac{1}{\sqrt{N!}}\sum_{P} sgn(P) \prod\phi_{P_j}(x_j),
\end{eqnarray}
is the anti-symmetric Slater determinant with $\phi_{i}(x)$ denoting the $i$-th single-particle wave-function and
\begin{eqnarray}
A=\prod_{1\leq j<l\leq N_b}sgn(x_{jb}-x_{lb})\prod_{j=1}^{N_b}\prod_{l=1}^{N_f}sgn(x_{jb}-x_{lf})
\end{eqnarray}
is a product of sign functions. The eigenenergy is given by $E=\sum_{l=1}^{N}\epsilon_{j_l}$ where $j_l$s are $N$ different integers and $\epsilon_{j_l}$ is the $j_l$-th single particle energy level. For convenience, we assume $\epsilon_{i}\leq \epsilon_{j}$ if $i<j$, and then the ground state energy is given by $E_0 =\sum_{l=1}^{N}\epsilon_{l}$. The ground state is highly degenerate, corresponding to different spin configurations.

When the system with weak p-wave interactions deviates from the TG limit, the degeneracy of the ground states is lifted and the energy can be expanded around its TG limit in terms of the small parameters $1/g_{bb}$,  $1/g_{bf}$ and $g_{ff}$:
\begin{eqnarray}
E=E_{TG}- \frac{1}{g_{bb}} I_{bb} -  \frac{1}{g_{bf}}I_{bf} -  g_{ff} I_{bf},
\end{eqnarray}
with
\begin{eqnarray*}
&&I_{bb}=\frac{N_b(N_b-1)}{2}\int dx dX\\\notag
&&\left| \frac{\partial\Psi(x_i,b;x_j,b;X)}{\partial r}|_{r=0^+}-\frac{\partial\Psi(x_i,b;x_j,b;X)}{\partial r}|_{r=0^-} \right|^2, \\\notag
&&I_{bf}=N_b N_f\int dx dX\\\notag
&&\left|\frac{\partial\Psi(x_i,b;x_j,f;X)}{\partial r}|_{r=0^+}-\frac{\partial\Psi(x_i,b;x_j,f;X)}{\partial r}|_{r=0^-}\right|^2, \\\notag
&&I_{ff}=2N_f(N_f-1)\int dx dX \left|\frac{\partial \Psi(x_i,f;x_j,f;X)}{\partial r}|_{r=0}\right|^2,
\end{eqnarray*}
where $\Psi(x_1,\sigma_1; x_2,\sigma_2...;x_N,\sigma_N)$ is the many-body eigen-wavefunction of the system, $x=(x_i+x_j)/2$, $r=x_i-x_j$ and $X$ denotes the remaining coordinates except $x_i$ and $x_j$.
Here $I_{bb}$, $I_{bf}$ and $I_{ff}$ are proportional to Tan's contacts \cite{Tan,Braaten,Barth}. Their expressions can be derived by using the Feymann-Hellmann theorem (see appendix for details).

\subsection{Variational perturbation calculation}
In the TG limit, the ground state degeneracy for the system composed of $N_b$ bosons and $N_f$ spinless fermions is $D=\frac{(N_b+N_f)!}{N_b! N_f!}$,  corresponding to different configurations of $N_b$ bosons in $N$ single-particle states. Once the system deviates from the infinite repulsion limit, the degeneracy of ground state manifold is lifted. As long as the system is still in the strongly interacting regime with $1/g_{bb}$, $1/g_{bf}$ and $g_{ff}$ much smaller than the single particle level space, we can use the degenerative perturbation to calculate the energy splitting.
Here we adopt the method developed in Ref.\cite{hu}. Since the particles can not penetrate each other in the TG limit, we can divide the real space into $N!$ subspaces. By introducing the step function
$$ \theta_{\alpha}=\left\{
\begin{aligned}
1 &   &(x_{\alpha_1}<x_{\alpha_2}<...<x_{\alpha_N}) \\
0 &   &(others) \\
\end{aligned}
\right.
$$
with $\alpha$ representing a sequence of $[1,2,...N]$, we can construct $N!$ orthogonal basis $\psi_A \theta_{\alpha}$. Considering constrains of the Bose-Fermi statistics, the number of allowed eigenfunctions is reduced to $D$. Using these constructed states as the basis of the degenerate space and defining permutation operators for bosons and fermions as $P_b$ and $P_f$, respectively, we then obtain $D$ normalized and orthogonal basis of the degenerate subspace as follows:
\begin{eqnarray}
\psi_{\alpha}(x_1,x_2,...,x_N)=\sqrt{D}\sum_{P_b,P_f}(-1)^{P_b} (P_b P_f \theta_{\alpha})\psi_A .
\end{eqnarray}
When $1/g_{bb}, ~ 1/g_{bf}, ~g_{ff} \rightarrow 0$, the eigenfunction should approach its TG limt smoothly and fall into the degenerate subspace. By projecting the state into the degenerate subspace with the help of the projection operator:
 $P_{deg}=\sum_{\alpha}|\psi_{\alpha}\rangle\langle\psi_{\alpha}|$,
we can expand the eigenfunction as
\begin{eqnarray}
\Psi(x_1,...,x_N)=\sum_{\alpha}a_{\alpha}\psi_{\alpha} ,
\end{eqnarray}
with $\sum_{\alpha}|a_{\alpha}^2|=1$.

In terms of the above variational wavefuntion, the Bose-Bose contact $I_{bb}$, Bose-Fermi contact $I_{bf}$ and Fermi-Fermi contact $I_{ff}$ can be represented as:
\begin{eqnarray}
I_{bb}=\sum_{\alpha,\alpha'}a_{\alpha}^* a_{\alpha'}J^{bb}_{\alpha,\alpha'}=\overrightarrow{a}J^{bb}\overrightarrow{a}', \\
I_{bf}=\sum_{\alpha,\alpha'}a_{\alpha}^* a_{\alpha'}J^{bf}_{\alpha,\alpha'}=\overrightarrow{a}J^{bf}\overrightarrow{a}',\\
I_{ff}=\sum_{\alpha,\alpha'}a_{\alpha}^* a_{\alpha'}J^{ff}_{\alpha,\alpha'}=\overrightarrow{a}J^{ff}\overrightarrow{a}',
\end{eqnarray}
where $\overrightarrow{a}=(a_1,a_2...a_D)^T$ and the reduced contact matrices for Bose-Bose, Bose-Fermi and Fermi-Fermi interaction are defined as
\begin{widetext}
\begin{eqnarray}
J^{bb}_{\alpha,\alpha'}&=&\frac{N_b(N_b-1)}{2}  \int (\frac{\partial\psi_{\alpha}}{\partial r}|_{r=0^+}-\frac{\partial\psi_{\alpha}}{\partial r}|_{r=0^-})^{*}(\frac{\partial\psi_{\alpha'}}{\partial r}|_{r=0^+}-\frac{\partial\psi_{\alpha'}}{\partial r}|_{r=0^-})dx dX ,\\
J^{bf}_{\alpha,\alpha'}&=& N_b N_f\int  (\frac{\partial\psi_{\alpha}}{\partial r}|_{r=0^+}-\frac{\partial\psi_{\alpha}}{\partial r}|_{r=0^-})^{*}(\frac{\partial\psi_{\alpha'}}{\partial r}|_{r=0^+}-\frac{\partial\psi_{\alpha'}}{\partial r}|_{r=0^-})dx dX ,\\
J^{ff}_{\alpha,\alpha'}&=& 2N_f(N_f-1)\int (\frac{\partial\psi_{\alpha}}{\partial r}|_{r=0})^{*}(\frac{\partial\psi_{\alpha'}}{\partial r}|_{r=0})dx dX .
\end{eqnarray}
\end{widetext}
Then the energy can be represented as
\begin{widetext}
\begin{eqnarray}
E=E_{TG}-\frac{1}{g_{bb}} \sum_{\alpha,\alpha'}a_{\alpha}^* a_{\alpha'}J^{bb}_{\alpha,\alpha'}- \frac{1}{g_{bf}} \sum_{\alpha,\alpha'}a_{\alpha}^* a_{\alpha'} J^{bf}_{\alpha,\alpha'}
-g_{ff}\sum_{\alpha,\alpha'}a_{\alpha}^* a_{\alpha'}J^{ff}_{\alpha,\alpha'}.
\end{eqnarray}
\end{widetext}
Introducing the anisotropy parameters $\gamma = \frac{g_{bb}}{g_{bf}}$ and $\gamma'={g_{bb}}{g_{ff}}$, the above equation can be simplified as
\begin{eqnarray}
E=E_{TG}-\frac{1}{g_{bb}} \sum_{\alpha,\alpha'}a_{\alpha}^* a_{\alpha'} J_{\alpha,\alpha'},
\end{eqnarray}
with the total contact defined as $J_{\alpha,\alpha'}=J^{bb}_{\alpha,\alpha'} + \gamma J^{bf}_{\alpha,\alpha'}+ \gamma' J^{ff}_{\alpha,\alpha'}$
We then determine the contact via the variational principle by minimizing $L=E-\lambda(\sum_{\alpha}a_{\alpha}^* a_{\alpha}-1)$ \cite{Guan-thesis,Zinner14,hu}, which leads to
\begin{eqnarray}
\sum_{\alpha'}J_{\alpha,\alpha'}a_{\alpha'}=\lambda a_{\alpha} . \label{eq-contact}
\end{eqnarray}
It is obvious that $\lambda$ and $\vec{a}$ are eigenvalue and eigenvector of the total $D\times D$ contact matrix $J$ \cite{Guan-thesis}. The diagonalization of the contact matrix directly gives the splitting energy of the degenerate levels in the TG limit:
\begin{eqnarray}
E= E_{TG}- \frac{\lambda}{g_{bb}} .
\end{eqnarray}

Next, we apply the variational perturbation method to study the ground state properties of the equal-mixing mixture composed of 2 bosons and 2 fermions in a harmonic trap with $1/g_{bb}\ll 1$,  $1/g_{bf}\ll 1$ and $g_{ff} \ll 1$. At TG limit, the ground state is $6$-fold degenerate. Denoting $x_1,x_2$ and $x_3,x_4$ as coordinates of bosons and fermions, respectively, then we can define the six distinct subspace basis fulfilling exchange statistics:
\begin{eqnarray*}
\psi_1=\sqrt{6}(\theta(1234)-\theta(2134)+\theta(1243)-\theta(2143))\psi_A,\\\notag
\psi_2=\sqrt{6}(\theta(1324)-\theta(2314)+\theta(1423)-\theta(2413))\psi_A,\\\notag
\psi_3=\sqrt{6}(\theta(1342)-\theta(2341)+\theta(1432)-\theta(2431))\psi_A,\\\notag
\psi_4=\sqrt{6}(\theta(3124)-\theta(3214)+\theta(4123)-\theta(4213))\psi_A,\\\notag
\psi_5=\sqrt{6}(\theta(3142)-\theta(3241)+\theta(4132)-\theta(4231))\psi_A,\\\notag
\psi_6=\sqrt{6}(\theta(3412)-\theta(3421)+\theta(4312)-\theta(4321))\psi_A,
\end{eqnarray*}
where $\psi_A$ represents the Slater determinant composed of the $N$ lowest eigen-functions (here $N=4$) given by $\Delta=C_N [\prod_{i=1}^{N}e^{-\xi_i^2/2}]\prod_{1\leq j<k\leq N}(\xi_k-\xi_j)$ with  coefficient $C_N=2^{N(N-1)/4}a_{\omega}^{-N/2}[N!\prod_{n=0}^{N-1}(n!\sqrt{\pi})]^{-1/2}$ and $\xi=x/a_{\omega}\equiv x/\sqrt{1/{\omega}}$.
After some straightforward calculations, we get the contact matrices for the boson-boson interaction and boson-fermion interaction,
\begin{eqnarray*}
J^{bb}=\frac{32}{3\sqrt{\pi}a_{\omega}^3}\left(
\begin{array}{cccccc}
A_1 & 0   & 0   &   0   &  0  &  0\\
0   & 0   & 0   &   0   &  0  &  0\\
0   & 0   & 0   &   0   &  0  &  0\\
0   & 0   & 0   &   A_2 &  0  &  0\\
0   & 0   & 0   &   0   &  0  &  0\\
0   & 0   & 0   &   0   &  0  &  A_1\\
\end{array}
\right)
\end{eqnarray*}
and
\begin{eqnarray*}
J^{bf} &=& \frac{16}{3\sqrt{\pi}a_{\omega}^3} \times \\
& & \left(
\begin{array}{cccccc}
A_2  & -A_2     & 0     &  0     &  0         &  0 \\
-A_2 & 2A_1+A_2 & -A_1  &  -A_1  &  0         &  0 \\
0    & -A_1     & 2A_1  &  0     &  -A_1      &  0 \\
0    & -A_1     & 0     &  2A_1  &  -A_1      &  0 \\
0    & 0        & -A_1  &  -A_1  &  2A_1+A_2  &  -A_2 \\
0    & 0        & 0     &  0     &  -A_2        &  A_2 \\
\end{array}
\right),
\end{eqnarray*}
respectively, and also the contact matrix for the fermion-fermion p-wave interaction,
\begin{eqnarray*}
J^{ff}=\frac{32}{3\sqrt{\pi}a_{\omega}^3}\left(
\begin{array}{cccccc}
A_1 & 0   & 0   &   0   &  0  &  0\\
0   & 0   & 0   &   0   &  0  &  0\\
0   & 0   & A_2   &   0   &  0  &  0\\
0   & 0   & 0   &   0 &  0  &  0\\
0   & 0   & 0   &   0   &  0  &  0\\
0   & 0   & 0   &   0   &  0  &  A_1\\
\end{array}
\right).
\end{eqnarray*}
Here $A_1=0.5938$ and $A_2=0.7796$ are integral constants. By solving Eq.(\ref{eq-contact}) for the given contact matrix, we can get the variational wavefunctions and energies.
\begin{figure}
\includegraphics[height=3.8in,width=3.8in]{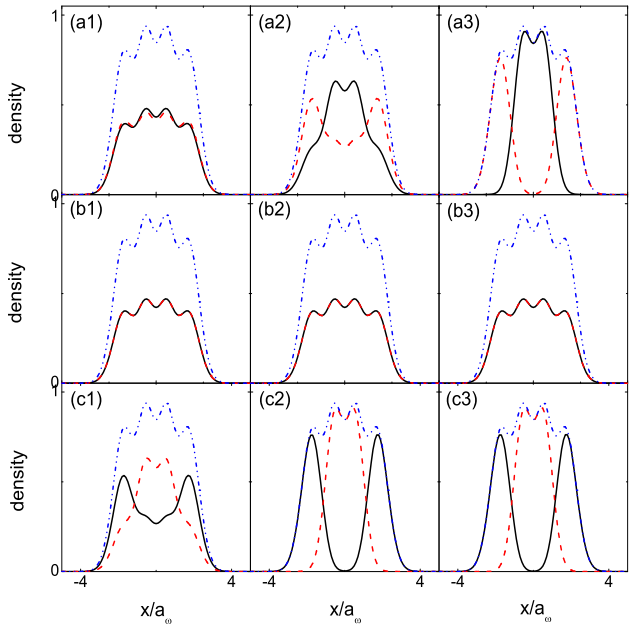}
\caption{(Color online) Density distributions for different $\gamma$ and $\gamma'$. From (a1) to (a3),$\gamma=10$, $\gamma=1$, $\gamma=0.1$ with $\gamma'=0.1$; from (b1) to (b3),$\gamma=10$, $\gamma=1$, $\gamma=0.1$ with $\gamma'=1$; from (c1) to (c3), $\gamma=10$, $\gamma=1$, $\gamma=0.1$ with $\gamma'=10$. The black solid, red dashed and blue dashed dotted lines denote boson, fermion and total density distributions, respectively.}
\end{figure}

In Fig.1, we demonstrate the density profiles of the Bose-Fermi mixtures composed of $2$ bosons and $2$ fermions for various cases with different anisotropic parameters $\gamma$ and $\gamma'$. In each row, we keep $\gamma'$ fixed but decrease $\gamma$ from left to right, corresponding to the increase of relative strength of boson-fermion interaction, whereas in each column, we keep $\gamma$ fixed but increase $\gamma'$ from top to bottom, corresponding to the increase of the p-wave interaction strength $g_{ff}$.  For cases of (a1)-(a3) with $\gamma' \ll 1$, the p-wave interaction is much weaker than the inverse of strongly repulsive boson-boson interaction strength, the density distributions resemble that of systems with only boson-boson interactions and boson-fermion interactions \cite{hu}. 
For a large $\gamma$ with $g_{bb} \gg g_{bf}$, the bosons behave like hard-core bosons exhibiting nearly the same distributions as that of fermions. As $\gamma$ decreases, the repelling between bosons and fermions increases, leading to that the fermions are repelled from the trap center. For cases of (b1)-(b3) with $\gamma'=1$, it is interesting to notice that the density distributions of bosons and fermions are identical and do not change with the variation of $\gamma$. These results can be understood from 
the existence of a Bose-Fermi duality between a 1D spinless fermionic system with p-wave interaction strength $g_{ff}$ and a bosonic interacting system with interaction strength $1/g_{ff}$ \cite{Cheon,Girardeau05,Girardeau06,Girardeau04,Granger}. Via this Bose-Fermi duality, the system with $\gamma'=1$ can be mapped to a two-component Bose-Bose mixture with equal intra-component interaction strengths, and thus the density distribution of each component should be the same according to the symmetry for interchanging different-component bosons.
For $\gamma'\gg 1$ as shown in (c1) to (c3), the p-wave interaction drastically influences the density distributions for fermions. The bosons exhibit stronger interaction and are repelled from the center. We can also understand these results from the Bose-Fermi duality as the spinless fermions can be mapped to another-component bosons with weaker intra-component interactions.

We should notice that though for $\gamma'=1$, the density distributions for bosons and fermions are the same, their momentum distributions are quite different, reflecting their different statistics, as shown in Fig.2. With the decrease of anisotropy $\gamma$, the momentum distributions for fermions become wider and wider and exhibit two peaks for small $\gamma$, while the momentum distribution for bosons always exhibits a peak at the zero momentum.
\begin{figure}
\includegraphics[width=3.7in]{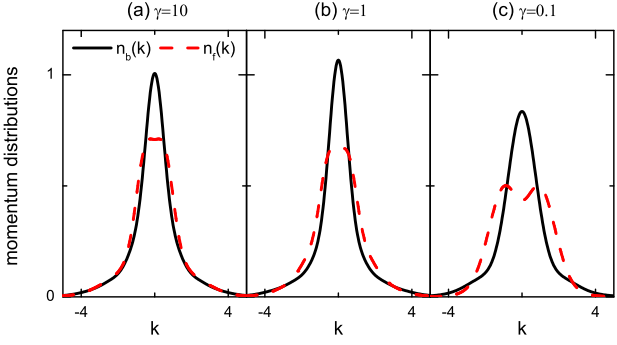}
\caption{(Color online) Momentum distributions for different $\gamma$ with $\gamma'=1$. The black solid, red dashed lines denote momentum distributions for bosons and fermions respectively.}
\end{figure}

\subsection{Effective spin model}
While the total density distributions are almost identical in the TG limit, their spin densities differ and are sensitive to the small parameters, as shown in our variational calculations.  Several recent theoretical works have demonstrated that the structure of the quasidegenerate ground-state multiplet of strongly interacting multi-component quantum gas can be determined by an effective spin-chain model \cite{PuHan,Deuretzbacher14,Zinner15,cuixiaoling}. It has also been shown that the spin distributions can be well reproduced by using the effective spin-exchange Hamiltonian.
Following the standard procedures \cite{PuHan,Deuretzbacher14,Zinner15,cuixiaoling}, we can get an effective spin model of the strongly interacting Bose-Fermi mixture in the presence of weak p-wave interactions, which is written as
\begin{eqnarray}
H_{eff}&=& -\sum_{m}I_m \left[ \frac{1}{g_{bb}} P_{m,m+1}^{bb}+\frac{1}{g_{bf}} P_{m,m+1}^{t} \right. \notag\\
& & \left.~~~~~~~~~~ +g_{ff}P_{m,m+1}^{ff} \right], ~
\end{eqnarray}
where $P_{m,m+1}^{bb}$, $P_{m,m+1}^{bf}$, $P_{m,i+1}^{t}$ are projection operators into the corresponding interacting channels, that is:
\begin{eqnarray*}
P_{m,m+1}^{bb}&=& |b_m b_{m+1}\rangle\langle b_m b_{m+1}| \\\notag
P_{m,m+1}^{ff}&=&|f_m f_{m+1}\rangle\langle f_m f_{m+1}| \\\notag
P_{m,m+1}^{t}&=&\frac{1}{2}(|b_m f_{m+1}\rangle+|f_m b_{m+1}\rangle) \times \notag\\
&& ~~~(\langle b_m f_{m+1}|+\langle f_m b_{m+1}|) .
\end{eqnarray*}
The $I_m$s are integral coefficients given by
\begin{equation}
I_m=2 N!\int \prod_{l} dx_l \delta(x_{m+1}-x_{m}) \theta_{[m+1,m]}\left|\frac{\partial \Psi_A}{\partial x_m}\right|^2
\end{equation}
with $ \theta_{[m+1,m]}=\theta_{[1,2,\cdots,N]}/\theta(x_{m+1}-x_{m})$ being a reduced sector function \cite{PuHan}.
In the spin space spanned by inner state of $m$ and $m+1$ particles (the four basis are $|bb\rangle$, $|bf\rangle$, $|fb\rangle$, $|ff\rangle$), the operators acting on this spin space can be expressed as the following matrix:
\begin{eqnarray}
H_{m,m+1}=I_m \left(
\begin{array}{cccc}
-\frac{1}{g_{bb}}      &       0     &     0       & 0\\
0              &   -\frac{1}{2g_{bf}} &  -\frac{1}{2g_{bf}}  & 0\\
0              &   -\frac{1}{2g_{bf}} &  -\frac{1}{2g_{bf}}  & 0\\
0              &     0       &    0        & -g_{ff} \\
\end{array}
\right)
\end{eqnarray}
The above matrix can be represented by spin operators as:
\begin{eqnarray}
H_{eff}&=& -\sum_m \frac{I_m}{g_{bb}} \left[ \gamma (S_m^x S_{m+1}^x+S_m^y S_{m+1}^y) \right. \notag \\
& & \left. + (2-\gamma) S_m^z S_{m+1}^z + ( \gamma' - 1)n_m^f n_{m+1}^f \right]
\end{eqnarray}
after neglecting a constant term. Here the spin operators are defined as $S_m^+=b_m^{\dag}f_m$, $S_m^z=(n_m^b-n_m^f)/2$, and the anisotropy parameters are defined as before.
Consider the case with $g_{ff}=1/g_{bb}$, i.e., $\gamma'=1$, the last term in the above Hamiltonian vanishes, we then get the following simple form for the effective spin model:
\begin{eqnarray}
H_{eff}&=& -\sum_m \frac{I_m}{ g_{bf}} \left[ (S_m^x S_{m+1}^x+S_m^y S_{m+1}^y)+\right. \notag\\
&& \left.(2/\gamma -1) S_m^z S_{m+1}^z \right]
\label{XXZ}
\end{eqnarray}
Based on this effective model, we can expect that a first order phase transition from AFM state to FM state occurs by tuning $\gamma$.

For a uniform system under periodic boundary conditions, all $I_m$s are equal (set as $I$), and the effective Hamiltonian reduces to a standard spin $XXZ$ model \cite{YangCN} $H =  -\sum_m  J [ (S_m^x S_{m+1}^x+S_m^y S_{m+1}^y)+ \Delta S_m^z S_{m+1}^z ]$ with $J=\frac{I}{g_{bf}}$ and the anisotropy parameter $\Delta=2/\gamma -1$. Since $J>0$ ($I>0$ and $g_{bf}>0$ ), we can get the conclusion that the system will undergo a first order transition from a critical AFM phase ( $\gamma>1$) to a ferromagnetic (FM) phase ( $\gamma<1$) by tuning the anisotropy $\gamma$. For systems in harmonic traps, as shown in Fig.3(a), we demonstrate the energy levels with respect to anisotropy $1/\gamma$ for the effective spin model of the system composed of four particles. The level crossing appearing at $\gamma=1$ indicates that a first-order phase transition occurs. From Fig.3(b), we can see clearly that it is a transition from a AFM phase to FM phase.
\begin{figure}
\includegraphics[width=3.7in]{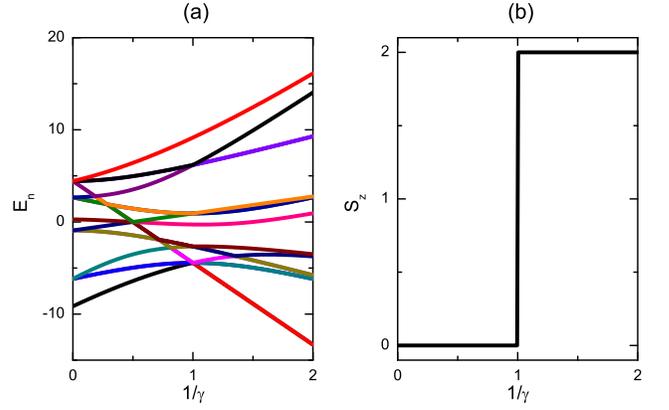}
\caption{(Color online) (b) Energy spectrum $E_n$ (in units of $\frac{1}{a_{\omega}^3 g_{bf}}$ with respect to $1/\gamma$ for $N=4$ particles in harmonic traps. (b) $z$ component magnetization with respect to $1/\gamma$.}
\end{figure}


\section{Spin-$1/2$ fermionic gas with weak p-wave interactions}
In this section, we consider the spin-$1/2$ fermionic gas with p-wave interactions described by the following dimensionless Hamiltonian
\begin{eqnarray}
H&=&\sum_{i} [-\frac{1}{2} \frac{\partial^2}{\partial x_{i}^2}+V(x_{i})] +  g_{\uparrow \downarrow} \sum_{i<j}\delta(x_{i}-x_{j})\delta_{\sigma_i,-\sigma_j}\notag\\
 &+&  g_{\uparrow \uparrow }\sum_{i<j}(\frac{\partial}{\partial x_i}-\frac{\partial}{\partial x_j})\delta(x_i-x_j)(\frac{\partial}{\partial x_i}-\frac{\partial}{\partial x_j})\delta^{\uparrow \uparrow}_{\sigma_i,\sigma_j} \notag\\
  &+& g_{\downarrow \downarrow}\sum_{i<j}(\frac{\partial}{\partial x_i}-\frac{\partial}{\partial x_j})\delta(x_i-x_j)(\frac{\partial}{\partial x_i}-\frac{\partial}{\partial x_j})\delta^{\downarrow \downarrow}_{\sigma_i,\sigma_j}, ~~~~ \label{H-FF}
\end{eqnarray}
where $g_{\uparrow \downarrow} $ is the dimensionless s-wave interaction parameter between inter-component fermions, and $g_{\uparrow \uparrow }$ and $g_{\downarrow \downarrow}$ are dimensionless p-wave interaction parameters between intra-component fermions. In the absence of p-wave interactions, the TG limit of the spin-$1/2$ fermionc gas has been studied in Ref. \cite{Girardeau07,Guan09}.

When the system with weak p-wave interactions deviates from the TG limit, the degeneracy of ground state is lifted and the energy can be expanded around its TG limit in terms of the small parameters $1/g_{\uparrow \downarrow}$, $g_{\uparrow \uparrow }$ and $g_{\downarrow \downarrow}$:
\begin{eqnarray}
E=E_{TG}- \frac{1}{g_{\uparrow \downarrow}} I_{\uparrow \downarrow} -  {g_{\uparrow \uparrow }}I_{\uparrow \uparrow } -  g_{\downarrow \downarrow} I_{\downarrow \downarrow},
\end{eqnarray}
where $I_{\uparrow \downarrow}$,
$I_{\uparrow \uparrow }$ and
$I_{\downarrow \downarrow}$ are proportional to Tan's contacts and can be calculated with the help of Feymann-Hellmann theorem. Explicitly, we have
\begin{eqnarray*}
&&I_{\uparrow\downarrow}=N_{\uparrow} N_{\downarrow}\int dx dX\\\notag
&&\left|\frac{\partial\Psi(x_i,\uparrow;x_j,\uparrow;X)}{\partial r}|_{r=0^+}-\frac{\partial\Psi(x_i,\uparrow;x_j,\downarrow;X)}{\partial r}|_{r=0^-}\right|^2\\\notag
&&I_{\uparrow\uparrow}=2N_{\uparrow}(N_{\uparrow}-1)\int \left|\frac{\partial \Psi(x_i,\uparrow;x_j,\uparrow;X)}{\partial r}|_{r=0}\right|^2dx dX\\\notag
&&I_{\downarrow\downarrow}=2N_{\downarrow}(N_{\downarrow}-1)\int\left|\frac{\partial \Psi(x_i,\downarrow;x_j,\downarrow;X)}{\partial r}|_{r=0}\right|^2dx dX .
\end{eqnarray*}
\begin{figure}
\includegraphics[width=3.7in]{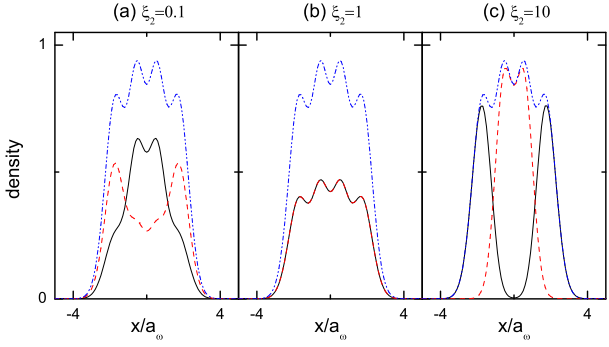}
\caption{(Color online) Density distribution of 2 spin-up fermions and 2 spin-down fermions with $\xi_1=1$. From (a) to (c), $\xi_2$=0.1, 1, 10. The black solid, red dashed and blue dashed dotted lines denote spin-up fermions, spin-down fermions and total distributions, respectively..}
\end{figure}
\begin{figure}
\includegraphics[width=3.7in]{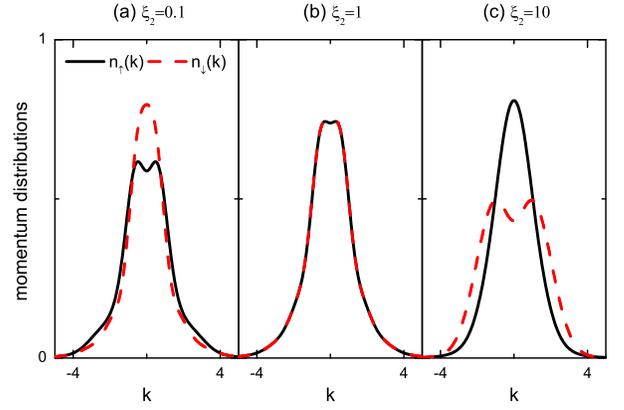}
\caption{(Color online) Momentum distributions of 2 spin-up fermions and 2 spin-down fermions with $\xi_1=1$. From (a) to (c), $\xi_2$=0.1, 1, 10. The black solid, red dashed lines denote momentum distributions for spin-up and spin-down fermions, respectively.}
\end{figure}

In the same framework of the variational perturbation theory for the Bose-Fermi mixture, we can calculate the variational ground state wavefunction of the Fermi-Fermi mixtures with weak p-wave interactions. Define $\xi_1=g_{\uparrow\downarrow}g_{\uparrow\uparrow}$ and $\xi_2=g_{\uparrow\downarrow}g_{\downarrow\downarrow}$. In Fig.4, we show the density profiles for $N_{\uparrow}=N_{\downarrow}=2$ systems for fixed $\xi_1=1$. For $\xi_2\ll 1$, the spin-up fermions will locate at the trap center while the spin-down fermions behave more like non-interacting fermions and are repelled to the trap wings. For $\xi_2\gg 1$, things will totally change. The spin-up particles are repelled from the trap center. For equal p-wave interacting strength $\xi_1=\xi_2$, the spin-up and spin-down fermions share the same distributions. In Fig.5, we demonstrate the momentum distributions with fixed $\xi_1=1$ in accordance with Fig.4. The momentum distributions show opposite behaviors for $\xi_1<1$ and $\xi_2>1$. In the former case, the distributions for spin-up fermions exhibit two main peaks while the spin-down fermions exhibit one main peak. For $\xi_1=\xi_2$, the momentum distributions for the two components are the same.

Similar to the Bose-Fermi mixture discussed in the above section, the structure of the quasidegenerate ground-state multiplet of strongly interacting spin-$1/2$ quantum gas with weak p-wave interaction can be also determined by an effective spin-chain model. The effective spin Hamiltonian for the case without p-wave interactions has been derived in several recent works \cite{PuHan,Deuretzbacher14,Zinner15,cuixiaoling}. We can derive the effective Hamiltonian for the case in the presence of p-wave interactions in the same framework.
In the spin space spanned by inner state of $m$ and $m+1$ particles (the four basis are $|\uparrow\uparrow\rangle$, $|\uparrow\downarrow\rangle$, $|\downarrow\uparrow\rangle$, $|\downarrow\downarrow\rangle$), the operators acting on this spin space can be expressed as the following matrix:
\begin{eqnarray}
H_{m,m+1}=I_m \left(
\begin{array}{cccc}
-g_{\uparrow\uparrow}      &       0     &     0       & 0\\
0              &   -\frac{1}{2g_{\uparrow\downarrow}} &  \frac{1}{2g_{\uparrow\downarrow}}  & 0\\
0              &   \frac{1}{2g_{\uparrow\downarrow}} &  -\frac{1}{2g_{\uparrow\downarrow}}  & 0\\
0              &     0       &    0        & -g_{\downarrow\downarrow} \\
\end{array}
\right),
\end{eqnarray}
In terms of the standard spin-$1/2$ representation of Pauli matrices,
\begin{eqnarray}
\vec{S}_m=\frac{1}{2}f_{m\sigma}^{\dag}\vec{\sigma}_{\sigma \sigma'}f_{m\sigma'},
\end{eqnarray}
the above matrix can be represented by the spin operators as:
\begin{eqnarray}
H_{eff}&=&\sum_i\frac{I_m}{g_{\uparrow\downarrow}}\left[(S_m^x S_{m+1}^x+S_m^y S_{m+1}^y)+ (1-2\xi_1) S_m^z S_{m+1}^z \right. \notag \\
&& \left. +(\xi_1-\xi_2)n_{m\downarrow} n_{m+1\downarrow} \right]
\end{eqnarray}
after neglecting a constant term. For the case without p-wave interactions,
we have $\xi_1=\xi_2=0$, and the effective model reduces back to an effective isotropic AFM Heisenberg model \cite{PuHan,Deuretzbacher14,Zinner15}.

For the general case with $\xi_1 \neq \xi_2$, the effective model can not be described by a pure spin model as the last term can not be canceled out. However, for the case with $g_{\uparrow\uparrow}=g_{\downarrow\downarrow}$, we have $\xi_1=\xi_2=\xi$, and the spin model can be further simplified as:
\begin{eqnarray}
H_{eff}= \sum_i\frac{I_m}{g_{\uparrow\downarrow}}[(S_m^x S_{m+1}^x+S_m^y S_{m+1}^y)+(1-2\xi) S_m^z S_{m+1}^z ]. \notag \\
\end{eqnarray}
Similar to the effective spin model discussed in the previous section, a phase transition from a critical AFM phase ( $\xi<1$) to a ferromagnetic (FM) phase ( $\xi>1$) can be induced by tuning p-wave interactions.

\section{Summary}
In summary, based on the variational perturbation theory within the degenerate ground state subspace in the TG limit, we have studied the one-dimensional strongly interacting Bose-Fermi mixtures and Fermi-Fermi mixtures with weak p-wave interactions between intra-component fermions. Our results demonstrate that the weak p-wave interactions play an important role in the regime slightly deviating from the TG limit with the strength of p-wave interactions being comparable with the inverse of the strength of repulsive s-wave interactions. The physical effects on density distributions can be understood by mapping the weak-interacting p-wave fermions to a strongly interacting bosons due to the existence of a Bose-Fermi duality.  In this parameter regime, we have also derived the effective spin models to describe the spin dynamics of strongly interacting gas mixtures.
Based on the effective spin-exchange models, we show that a phase transition from AFM phase to FM phase can be induced by tuning the anisotropic parameter of interactions between different components.
\begin{acknowledgments}
We thank L. Guan, Y. Jiang and X. Cui for helpful discussions. The work is supported by NSFC under Grants No. 11425419, No. 11374354 and No. 11174360.
\end{acknowledgments}

\appendix
\section{Derivation of the universal energy relations}
In this appendix, we give a clear derivation of the universal energy relations for Bose-Fermi mixtures with p-wave interactions.
Let $\Psi(x_1; x_2,...;x_N)$ be the normalized eigenstate of the system, which fulfills the Schr\"odinger equation:
\begin{eqnarray*}
H\Psi(x_1; x_2,...;x_N)=E \Psi(x_1; x_2,...;x_N) .
\end{eqnarray*}
Denote $x_i,x_j$ as the coordinates of two interacting particle with interaction $g_{bb}$ or $g_{bf}$ or $g_{ff}$, depending on the species. The coordinates of the remaining particles are denoted by $X$. In terms of center-of-mass and relative coordinates, $x=(x_i+x_j)/2$ and $r=x_i-x_j$, the p-wave interaction is equivalent to the following boundary conditions between two interacting fermions:
\begin{eqnarray*}
&&\frac{\partial\Psi(x_i,f;x_j,f;X)}{\partial r}|_{r=0^+}=\frac{\partial\Psi(x_i,f;x_j,f;X)}{\partial r}|_{r=0^-}\notag,\\
&&\Psi(x_i,f;x_j,f;X)|_{r=0^+}-\Psi(x_i,f;x_j,f;X)|_{r=0^-}\notag \\
&&=4 g_{ff}\frac{\partial \Psi(x_i,f;x_j,f;X)}{\partial r}|_{r=0},
\end{eqnarray*}
where we have used the condition that $\frac{\partial}{\partial r}|_{r=0^+}=\frac{\partial}{\partial r}|_{r=0^-}$ and defined $\frac{\partial}{\partial r}|_{r=0}=[\frac{\partial}{\partial r}|_{r=0^+}+\frac{\partial}{\partial r}|_{r=0^-}]/2$ for two interacting fermions.
For the boson-boson interaction, we have the following boundary condition:
\begin{eqnarray*}
&&\frac{\partial\Psi(x_i,b;x_j,b;X)}{\partial r}|_{r=0^+}-\frac{\partial\Psi(x_i,b;x_j,b;X)}{\partial r}|_{r=0^-}\notag\\
&&=g_{bb}\Psi(x_i,b;x_j,b;X)|_{r=0} .
\end{eqnarray*}
Similarly, for the boson-fermion interaction, we have
\begin{eqnarray*}
&&\frac{\partial\Psi(x_i,b;x_j,f;X)}{\partial r}|_{r=0^+}-\frac{\partial\Psi(x_i,b;x_j,f;X)}{\partial r}|_{r=0^-}\notag \\
&&=g_{bf}\Psi(x_i,b;x_j,f;X)|_{r=0} .
\end{eqnarray*}

Using Feymann-Hellmann theorem, we have
\begin{widetext}
\begin{eqnarray*}
\frac{\partial E}{\partial (-1/g_{bb})}&=& g_{bb}^2 \frac{\partial E}{\partial g_{bb}} = g_{bb}^2\int\sum_{i<j}\delta(x_i-x_j)\delta^b_{\sigma_i,\sigma_j}|\Psi|^2dx_1 dx_2...dx_N\\\notag
&=&\frac{N_b(N_b-1)}{2}\int dx dX\left|\frac{\partial\Psi(x_i,b;x_j,b;X)}{\partial r}|_{r=0^+}-\frac{\partial\Psi(x_i,b;x_j,b;X)}{\partial r}|_{r=0^-}\right|^2.
\end{eqnarray*}
\end{widetext}
Similarly, we have
\begin{widetext}
\begin{eqnarray*}
\frac{\partial E}{\partial (-1/g_{bf})} &=& g_{bf}^2 \frac{\partial E}{\partial g_{bf}} = g_{bf}^2\int\sum_{i<j}\delta(x_i-x_j)\delta_{\sigma_i,-\sigma_j}|\Psi|^2dx_1 dx_2...dx_N\\\notag
&=&N_b N_f\int dx dX \left|\frac{\partial\Psi(x_i,b;x_j,f;X)}{\partial r}|_{r=0^+}-\frac{\partial\Psi(x_i,b;x_j,f;X)}{\partial r}|_{r=0^-}\right|^2 .
\end{eqnarray*}
\end{widetext}
For the p-wave interaction between fermions, we have
\begin{widetext}
\begin{eqnarray*}
\frac{\partial E}{\partial g_{ff}} &=& \int\sum_{i<j}|(\frac{\partial}{\partial x_i}-\frac{\partial}{\partial x_j})\Psi|^2\delta(x_i-x_j)\delta^f_{\sigma_i,\sigma_j} dx_1 dx_2...dx_N\\\notag
&=&2N_f(N_f-1)\int dx dX \left|\frac{\partial \Psi(x_i,f;x_j,f;X)}{\partial r}|_{r=0}\right|^2 .
\end{eqnarray*}
\end{widetext}
Now we define the following three kinds of contact
\begin{widetext}
\begin{eqnarray*}
I_{bb} &=& \frac{N_b(N_b-1)}{2}\int dx dX \left|\frac{\partial\Psi(x_i,b;x_j,b;X)}{\partial r}|_{r=0^+}-\frac{\partial\Psi(x_i,b;x_j,b;X)}{\partial r}|_{r=0^-}\right|^2,\\\notag
I_{bf} &=& N_b N_f\int dx dX \left|\frac{\partial\Psi(x_i,b;x_j,f;X)}{\partial r}|_{r=0^+}-\frac{\partial\Psi(x_i,b;x_j,f;X)}{\partial r}|_{r=0^-}\right|^2,\\\notag
I_{ff} &=& 2N_f(N_f-1)\int dx dX \left|\frac{\partial \Psi(x_i,f;x_j,f;X)}{\partial r}|_{r=0}\right|^2,
\end{eqnarray*}
\end{widetext}
then we get the following form for the differential energy
\begin{eqnarray}
{d E} = {- d (1/g_{bb})} I_{bb} - {d (1/g_{bf})} I_{bf} - {d (g_{ff})} I_{ff} .
\label{energy}
\end{eqnarray}

Similarly, for two-component fermionic systems with $N_{\uparrow}$ spin-up  and $N_{\downarrow}$ spin-down fermions, the universal energy relation is given by
\begin{eqnarray}
{d E} = - {d (1/g_{\uparrow\downarrow})} I_{\uparrow\downarrow}- {d (g_{\uparrow\uparrow})} I_{\uparrow\uparrow} -{ (g_{\downarrow\downarrow})} I_{\downarrow\downarrow},
\label{energy}
\end{eqnarray}
where
\begin{widetext}
\begin{eqnarray*}
I_{\uparrow\downarrow} &=& N_{\uparrow} N_{\downarrow}\int dx dX \left|\frac{\partial\Psi(x_i,\uparrow;x_j,\uparrow;X)}{\partial r}|_{r=0^+}-\frac{\partial\Psi(x_i,\uparrow;x_j,\downarrow;X)}{\partial r}|_{r=0^-}\right|^2,\\\notag
I_{\uparrow\uparrow} &=& 2N_{\uparrow}(N_{\uparrow}-1)\int\left|\frac{\partial \Psi(x_i,\uparrow;x_j,\uparrow;X)}{\partial r}|_{r=0}\right|^2dx dX,\\\notag
I_{\downarrow\downarrow} &=& 2N_{\downarrow}(N_{\downarrow}-1)\int\left|\frac{\partial \Psi(x_i,\downarrow;x_j,\downarrow;X)}{\partial r}|_{r=0}\right|^2dx dX.
\end{eqnarray*}
\end{widetext}

\end{document}